\documentclass[letterpaper]{article} 
\usepackage{aaai25}  
\usepackage{times}  
\usepackage{helvet}  
\usepackage{courier}  
\usepackage[hyphens]{url}  
\usepackage{graphicx} 
\usepackage{natbib}  
\usepackage{caption} 
\usepackage{booktabs}
\usepackage{amsmath}
\usepackage{multirow}
\usepackage{pifont}
\newcommand{\checkmark}{\text{\ding{51}}}
\newcommand{\xmark}{\text{\ding{55}}}
\usepackage{algorithm}
\usepackage{algorithmic}

\usepackage{newfloat}
\usepackage{listings}
\DeclareCaptionStyle{ruled}{labelfont=normalfont,labelsep=colon,strut=off} 
\floatstyle{ruled}
\newfloat{listing}{tb}{lst}{}
\floatname{listing}{Listing}
\title{Complex-Cycle-Consistent Diffusion Model for Monaural Speech Enhancement}
\author{
    Yi Li\textsuperscript{\rm 1},
    Yang Sun\textsuperscript{\rm 2},
    Plamen Angelov\textsuperscript{\rm 1}
}
\affiliations{
    \textsuperscript{\rm 1}School of Computing and Communications, Lancaster University, UK\\

    \textsuperscript{\rm 2}Big Data Institute, University of Oxford, UK
}

\usepackage{bibentry}

\begin{document}

\maketitle

\begin{abstract}
In this paper, we present a novel diffusion model-based monaural speech enhancement method. Our approach incorporates the separate estimation of speech spectra's magnitude and phase in two diffusion networks. Throughout the diffusion process, noise clips from real-world noise interferences are added gradually to the clean speech spectra and a noise-aware reverse process is proposed to learn how to generate both clean speech spectra and noise spectra. Furthermore, to fully leverage the intrinsic relationship between magnitude and phase, we introduce a complex-cycle-consistent (CCC) mechanism that uses the estimated magnitude to map the phase, and vice versa. We implement this algorithm within a phase-aware speech enhancement diffusion model (SEDM). We conduct extensive experiments on public datasets to demonstrate the effectiveness of our method, highlighting the significant benefits of exploiting the intrinsic relationship between phase and magnitude information to enhance speech. The comparison to conventional diffusion models demonstrates the superiority of SEDM.
\end{abstract}

%

\section{Introduction}
In real-world acoustic environments, speech signals are inevitably contaminated by background noise, which can significantly deteriorate speech quality and intelligibility. The primary goal of speech enhancement techniques is to separate the target speech signal from the background noise. Consequently, speech enhancement plays a pivotal role in various speech processing systems, including assisted living, teleconferencing, and automatic speech recognition (ASR) \cite{seasr, seasr1}. Monaural speech enhancement presents one of the most challenging scenarios in this field, as it deals with a single channel.

Traditional deep learning-based methods for solving the monaural speech enhancement problem have been extensively studied. Recently, the diffusion model has not only achieved significant success in the field of image processing \cite{CVPR1, CVPR2} but has also been introduced into the field of speech enhancement with excellent results \cite{sedi, CDiffuse}. However, these methods have two limitations.

Firstly, these methods typically operate in the time-frequency (T-F) domain, where they estimate the magnitude response while leaving the phase response unaltered, as seen in \cite{IRM1, TAI}. However, Wang et al. introduced a novel approach by proposing a complex ideal ratio mask that allows for simultaneous enhancement of both magnitude and phase spectra, operating in the complex domain \cite{comp}. Recent studies, such as \cite{GN}, have widely adopted complex spectra in monaural speech enhancement due to their efficient performance improvement through the utilization of phase information from speech signals. Nevertheless, in most cases, magnitude and phase spectra are simultaneously enhanced by one or two neural networks to generate the final estimated speech spectra, neglecting the intrinsic relationship between magnitude and phase, which has been shown to be beneficial for further improving speech enhancement performance, as indicated in \cite{phase}. 

Secondly, diffusion models for speech enhancement \cite{sedi, CDiffuse} exploit Gaussian noise to learn denoising noisy speech based on maximum entropy and statistical inference. However, some recent diffusion-based image denoising techniques replace the Gaussian noise to real-world noise. Wu et al. synthesise realistic noise with the environmental settings to better model noise distribution complexity \cite{realnoise}. The experiments prove that using real-world noise can boost the performance rather than Gaussian noise. Nevertheless, there has been a lack of recent research attempting to substitute real-world noise for Gaussian noise in the filed of speech enhancement. This has sparked our interest in exploring the problem.

\begin{figure*}[htbp!]
\centering
\includegraphics[width=16.5cm, height=7.8cm]{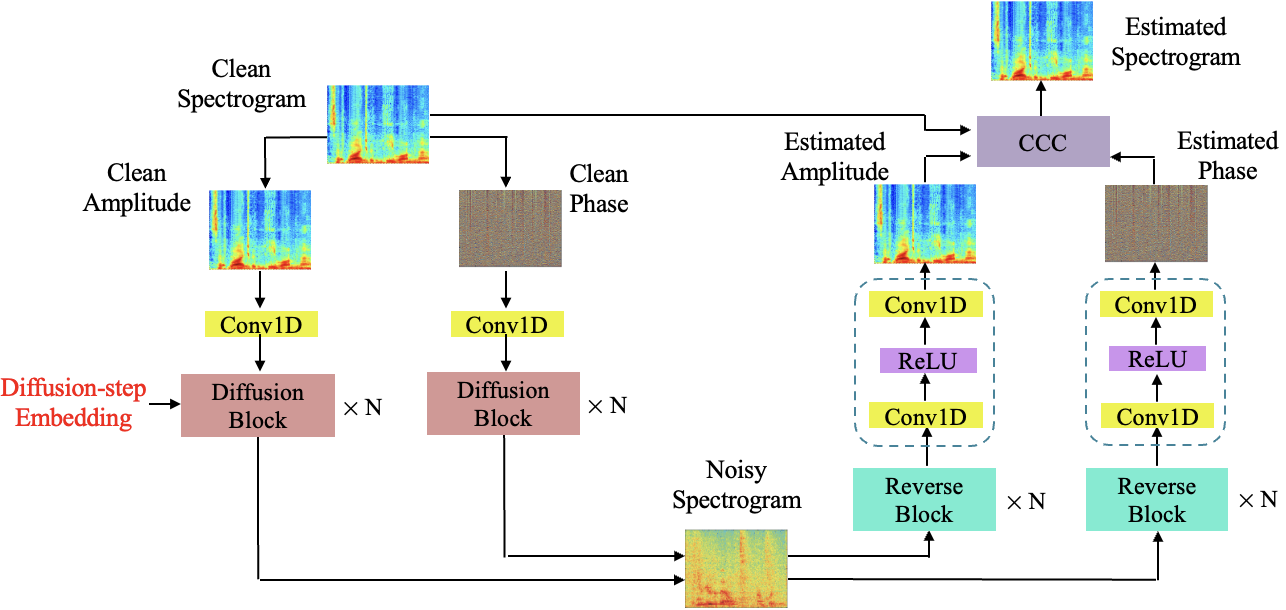}
\caption{The overall architecture of SEDM consisted of a diffusion process (left) and a noise-aware reverse process (right).}\centering
\label{overallstructure}
\end{figure*}

We present three contributions to address these limitations in this paper. Firstly, we propose a novel speech enhancement diffusion model (SEDM). In contrast to conventional diffusion model-based speech enhancement methods \cite{DM}, we replace the Gaussian noise in the forward process with real-world noise. In each embedding, a noise clip is randomly selected from different noise types, e.g., $dwashing$, $dliving$, and $pstation$ \cite{Demand}. Secondly, we propose a noise-aware reverse process that facilitates learning to generate both clean speech and noise spectra. Thirdly, we investigate the speech enhancement performance with independent networks. In encoders of SEDM, we independently estimate the noise distributions in the magnitude and phase of the mixture spectrogram. Subsequently, the decoders generate the estimated magnitude and phase of the target speech spectrogram. The relationship between magnitude and phase, as shown in \cite{phase}, is beneficial for further enhancing speech enhancement performance. Therefore, we calculate the commutative losses between the magnitude and phase features to further improve estimation accuracy.

\section{Related Works}
\subsection{Phase-Aware Speech Enhancement}
In conventional speech enhancement studies, the desired speech signal is typically reconstructed using the magnitude of the estimated speech signal and the phase of the noisy mixture. However, due to the reliance on noisy phase information, the speech enhancement performance may be compromised. Therefore, recent studies have started to incorporate the phase information of the desired speech signal. Polar coordinates (i.e. magnitude and phase) are commonly used when enhancing the STFT of noisy speech, as defined in (1):
\begin{equation}
S_{t, f}=\left|S_{t, f}\right| e^{i \theta_{S_{t, f}}}
\end{equation}
where $\left|S_{t, f}\right|$ represents the magnitude response and $\theta_{S_{t, f}}$ represents the phase response of the short-time Fourier transform (STFT) at time $t$ and frequency $f$. Each T-F unit in the STFT representation is a complex number with real and imaginary components.

\subsection{Diffusion Model}
Diffusion models, initially proposed in \cite{DM0}, have demonstrated strong generative capabilities. A typical diffusion probabilistic model comprises two key processes: a forward/diffusion process and a reverse process. In the forward process, the model transforms clean input data into an isotropic Gaussian distribution by introducing Gaussian noise to the original signal at each step. Conversely, in the reverse process, the diffusion probabilistic model predicts a noise signal and subtracts this predicted noise signal from the noisy input to recover the clean signal. As the first diffusion model-based speech enhancement work, Lu et al. introduced the concept of a supportive reverse process in their work \cite{DM}, where noisy speech is added at each time-step to the predicted speech signal. 



\section{Phase-Aware Diffusion Models}
The speech enhancement algorithm aims to estimate the speech signal $s(t)$ from noisy speech signal $y(t)$. To achieve that, we design a diffusion model-based training pipeline as presented in Figure 1. The proposed SEDM consists of a diffusion network (left side of Figure 1) and a noise-aware reverse network (left side of Figure 1) for a diffusion process and a reverse diffusion, respectively. In Figure. 1, the input and output are spectra, as we omit Short-Time Fourier Transform (STFT) for speech signals.
\subsection{Diffusion Process}
We disassemble STFT of clean speech signals into magnitude and phase components as the input for the diffusion network. The diffusion network consists of a stack of $N$ diffusion blocks, each with a residual channel size of $C$, followed by a 1$\times$1 convolutional layer. A single diffusion block is presented in Figure 2 (a).
\begin{figure}[htbp!]
\centering
\includegraphics[width=8cm, height=4.4cm]{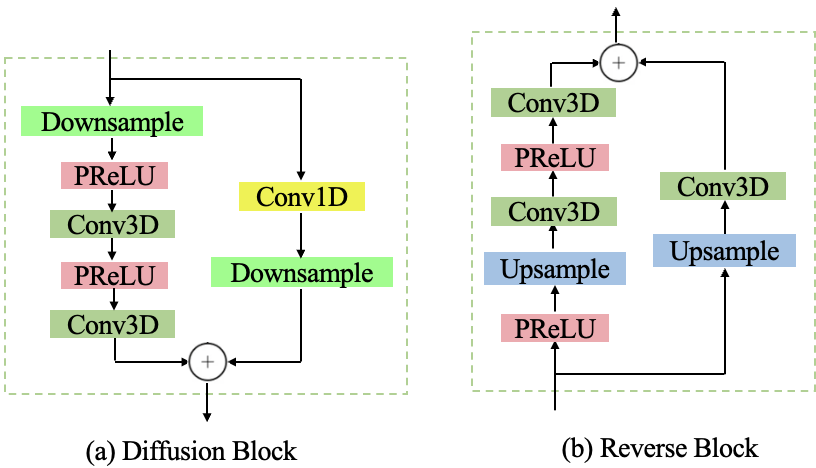}
\caption{The proposed diffusion block and reverse block.}\centering
\end{figure}

Similar to U-net models \cite{tf3, ushaped}, we replace pooling operations in each diffusion block with downsampling operators to complement a typical contracting network with successive layers. Additionally, we employ Parametric Rectified Linear Unit (PReLU) activation functions \cite{prelu}, which allow for flexibility in the range of values in the feature map, encompassing both negative and non-negative values. This flexibility contributes to more accurate estimation of the speech source. Furthermore, we incorporate residual connections \cite{rc} between the input of each diffusion block and the output of the final 3$\times$1 convolutional layer. This addition aids in achieving rapid convergence, particularly when training a deep diffusion network with a large value of $N$, by mitigating issues related to vanishing gradients. 

In addition, each diffusion block includes a skip connection \cite{rc} to the next block to maintain the desired information within the signal. Within the diffusion network, each diffusion block generates feature maps at specific resolutions, which are then scaled to produce a latent representation of the noisy speech feature with multiple resolutions. During the training of the diffusion network, the optimal weighted combinations of these multi-resolution spectra are learned in relation to the target, which is the original speech feature representation.

Within the forward process $q\left(Y_1, \cdots, Y_N \mid Y_0\right)$ of each diffusion block, we embed either the magnitude or phase of a noise clip's spectrogram into the input speech spectrogram $Y_0$ as:
\begin{equation}
q\left(Y_1, \cdots, Y_N \mid Y_0\right)=\prod_{n=1}^N q\left(Y_n \mid Y_{n-1}\right) 
\end{equation}
where $n$ is the index of diffusion block. Different from conventional diffusion models, we replace the Gaussian noise by real-world noises during the diffusion process. To achieve that, we divide four-minute noise sequences among all noise types from the DEMAND dataset \cite{Demand} into clips to align with the length of the input speech signals. We randomly select one clip $I_n$ and progressively add to a clean speech spectrogram as a stochastic differential equation (SDE) \cite{sde}:
\begin{equation}
d Y_n=\mu\left(Y_n, n\right) d n+\sigma\left(Y_n, n\right) d I_n
\end{equation}
where $\mu(Y_n, n)$ is the drift term representing the deterministic component and $\sigma(Y_n, n)$ is the diffusion term representing the stochastic (random) component. The final diffusion block generates the latent representation of the noisy speech spectra. 
 
\subsection{Reverse Process}
Different from conventional diffusion models, we propose a noise-aware reverse process that initiates the sampling process from the noisy speech spectrogram. At each reverse block, we estimate both the clean speech and noise spectra, all while minimizing the introduction of additional noise signals. A single reverse block is presented in Figure 2 (b).

In the proposed noise-aware reverse process, we aim to generate each $Y_{m-1}$ from the previous step $Y_{m}$. To achieve that, we define each reverse step with two trainable parameters $\gamma_m$ and $\theta_m$ as:
\begin{equation}
Y_{m-1}=\frac{1}{\gamma_m}\left(Y_m-\frac{\theta_m}{1-\gamma_m} I_m\right)+\sigma_m
\end{equation}
where $\sigma_m$ is the variance of the estimated speech spectra distribution, which can be calculated as:
\begin{equation}
\sigma_m = \frac{1-\bar{\gamma}_{m-1}}{1-\bar{\gamma}_m} \theta_m
\end{equation}

Again, similarly to the U-net models, reverse blocks utilize upsampling operators to restore the original spectrogram size. The final estimated magnitude and phase of the target speech spectrogram are derived from the last reverse blocks. Subsequently, the proposed CCC block further enhances the estimation accuracy of both magnitude and phase using the clean spectrogram.

Finally, the phase is reconstructed by re-wrapping the estimated unwrapped phase of the speech signal. This phase, along with the recovered speech magnitude, is used in the speech recovery module to reconstruct the estimated speech signal. During the test stage, the diffusion network is disregarded, and the reverse network is employed to enhance the noisy speech signals.
\section{Complex-Cycle-Consistent Learning}
After estimating the magnitude and phase of speech sources from the reverse blocks, they are input into the proposed Complex-Cycle-Consistent block (CCC) along with the clean magnitude and phase as shown in Figure 3. The block consists of two long short-term memory (LSTM) networks, with their parameters denoted as $\theta_A$ and $\theta_P$. We denote the estimated magnitude and phase of speech spectra as $\mathbf{S}_{A}$ and $\mathbf{S}_{P}$, respectively.

\begin{figure}[htbp!]
\centering
\includegraphics[width=7.3cm, height=9.2cm]{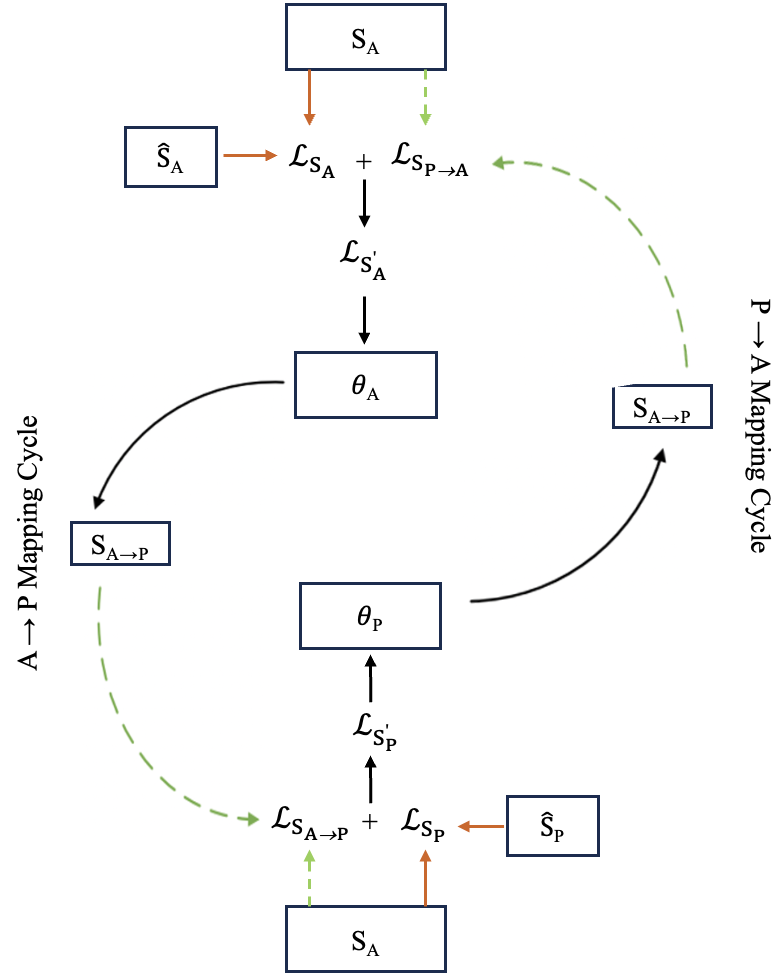}
\caption{The proposed complex-cycle-consistent learning for speech (CCC) enhancement.}\centering
\end{figure}

As the input, the magnitude and phase of the estimated speech are fed into the CCC module with the clean magnitude and phase. First, the magnitude loss is estimated with (3). We define $P\!\rightarrow\!A$ as the mapping of the spectra from the phase to the magnitude. We refer to $\mathbf{S}_{P \rightarrow A}$ as the new phase reconstruction from the last cycle. Then, the loss between the magnitude of clean speech spectra and $\mathbf{S}_{P \rightarrow A}$ is computed with the L2 norm of the error as:
\begin{equation}
\mathcal{L}_{\mathbf{S}_{P \rightarrow A}}= \|\mathbf{S}_{A}-\mathbf{S}_{P \rightarrow A}\|_{2}^{2}
\end{equation}
In the training stage, the loss term $\mathcal{L}_{\mathbf{S}_{P \rightarrow A}}$ is relatively large, in contrast to the loss $\mathcal{L}_{\mathbf{S}_{A}}$. Therefore, we use a weight $\lambda_{1}$ to attenuate $\mathcal{L}_{\mathbf{S}_{P \rightarrow A}}$. The combined magnitude loss can be obtained as:
\begin{equation}
\mathcal{L}_{\mathbf{S}_{A}^{\prime}} = \mathcal{L}_{\mathbf{S}_{A}} + \lambda_{1}\cdot\mathcal{L}_{\mathbf{S}_{P \rightarrow A}}
\end{equation}
The combined magnitude loss $\mathcal{L}_{\mathbf{S}_{A}^{\prime}}$ is applied to train $\theta_{A}$ which is used to map the spectra from the magnitude to the phase $\mathbf{S}_{A \rightarrow P}$ in the $A \rightarrow P$ mapping cycle. Similarly, the loss between the phase of the clean speech spectra and the reconstruction from the mapping $\mathbf{S}_{A \rightarrow P}$ is written as:
\begin{equation}
\mathcal{L}_{\mathbf{S}_{A \rightarrow P}}= \|\mathbf{S}_{P}-\mathbf{S}_{A \rightarrow P}\|_{2}^{2}
\end{equation}
Accordingly, the combined phase loss is presented as:
\begin{equation}
\mathcal{L}_{\mathbf{S}_{P}^{\prime}} = \mathcal{L}_{\mathbf{S}_{P}} + \lambda_{2}\cdot\mathcal{L}_{\mathbf{S}_{A \rightarrow P}}
\end{equation}
where a weight $\lambda_{2}$ is a weight parameter. Then, $\theta_{P}$ is trained with the combined phase loss and yields a new mapping magnitude reconstructions as $\mathbf{S}_{P \rightarrow A}$ for the next epoch. Parameters $\theta_A$ and $\theta_P$ trained with the cycle-consistent learning approach and finally outputs the magnitude and phase of estimated speech spectra. The pseudocode of the proposed CCC module is summarized as Algorithm 1. 

\begin{algorithm}
   \caption{Proposed complex-cycle-consistent learning}
   \label{alg:example}
\begin{algorithmic}[1]
   \STATE {\bfseries Input:} magnitude of the clean speech spectra $\mathbf{S}_{A}$, phase of the clean speech spectra $\mathbf{S}_{P}$, epoch $E = 1, 2, ...,$  $E_{max}$
   \STATE {\bfseries Output:} Estimated speech $\hat{\mathbf{S}}_{A}$ and $\hat{\mathbf{S}}_{P}$
   \STATE Initialize reverse network parameters $\theta_{A}$ and $\theta_{P}$
   \WHILE{$E = 1$}
  \STATE Estimate $\hat{\mathbf{S}}_{A}$ and $\hat{\mathbf{S}}_{P}$
  \STATE Calculate the losses: $\mathcal{L}_{\mathbf{S}_{A}}$ and $\mathcal{L}_{\mathbf{S}_{P}}$
  \ENDWHILE
   \FOR{$E = 2, ...,$  $E_{max}$}
   \STATE Run the mapping cycle $\mathbf{S}_{A \rightarrow P}$ by using $\theta_{A}$
   \STATE Update $\mathcal{L}_{\mathbf{S}_{A}^{\prime}}$ as (8)
   \STATE Run the mapping cycle $\mathbf{S}_{P \rightarrow A}$ by using $\theta_{P}$
   \STATE Update $\mathcal{L}_{\mathbf{S}_{P}^{\prime}}$ as (10)
   \STATE $\mathcal{L}_{\mathbf{S}_{A}}$ = $\mathcal{L}_{\mathbf{S}_{A}^{\prime}}$, $\mathcal{L}_{\mathbf{S}_{P}}$ = $\mathcal{L}_{\mathbf{S}_{P}^{\prime}}$
   \STATE Update $\theta_{A}$, $\theta_{P}$ by minimizing $\mathcal{L}_{\mathbf{S}_{A}}$ and $\mathcal{L}_{\mathbf{S}_{P}}$
   \ENDFOR
   \STATE Estimate $\hat{\mathbf{S}}_{A}$ and $\hat{\mathbf{S}}_{P}$ with trained $\theta_{A}$ and $\theta_{P}$

\end{algorithmic}
\end{algorithm}
\section{Experiments}
\begin{table*}[htbp!]
\centering
\caption{Speech enhancement performance comparisons on the \textbf{IEEE} and \textbf{TM} datasets. The number of residual blocks, channels and kernel is denoted as N, C, and K, respectively.}
\begin{tabular}{c|ccc|ccc|ccc}
\hline
 & \multicolumn{3}{c|}{Configuration}& \multicolumn{3}{c|}{IEEE }& \multicolumn{3}{c}{TIMIT}\\
\hline
Method & N & C & K & STOI ($\%$)  & PESQ  & fwSNRseg (dB)& STOI ($\%$)  & PESQ  & fwSNRseg (dB)\\    
 \hline
Unprocessed & - & - & - & 42.3&1.52 &3.11 & 41.5&1.44 &3.04\\
DCTCRN & 7 & 256 & 5  &73.4&2.36 &12.88 & 78.5&2.45 &13.24\\
RemixIT & 64 & 512 & 21 & 74.8&2.42 &13.13 & 79.3&2.56 &13.97\\
FRCRN  & 6 & 128 & 7 &76.5 & 2.50 & 13.87&80.2 & 2.59 & 14.43\\
CMGAN & 16 & 64 & 8 &75.2 & 2.47 & 12.98&79.6 & 2.55 & 13.78\\
SCP-GAN & - & - & - & 77.3 & 2.66 & 14.04&81.5 & 2.77 & 15.00\\
 \hline 
 \textit{SEDM-S} & 30 & 63 & 3 & 79.0 & 2.68 &  13.97& 81.2 & 2.76 &  14.85\\
  \textit{SEDM-M} & 40 & 128 & 3 & \underline{79.7} & \underline{2.73} &  \underline{14.22}& \underline{81.7} & \underline{2.81} &  \underline{15.18}\\
  \textit{SEDM-L} & 50 & 128 & 3 &{\bfseries 80.2} &{\bfseries 2.75} & {\bfseries 14.30}&{\bfseries 81.9} &{\bfseries 2.83} & {\bfseries 15.29}\\
 \hline 
\end{tabular}
\end{table*}

\subsection{Datasets}  \label{5.1}
We extensively perform experiments on several public speech datasets, including IEEE \cite{IEEE}, 	TIMIT Acoustic-Phonetic Continuous Speech Corpus (TIMIT) \cite{TM}, VOICE BANK (VCTK) \cite{VB}, and Deep Noise Suppression (DNS) challenge \cite{DNS}. To generate noisy speech signals in training and test, we randomly collect and use 10 of 15 noise types $psquare$, $dliving$, $dkitchen$, $nriver$, $tcar$, $dwashing$, $npark$, $omeeting$, $ohallway$ and $pstation$ from Diverse Environments Multichannel Acoustic Noise Database (DEMAND) \cite{Demand}. Each noise interference has a unique case and lasts four minutes long, and it is divided into two clips with an equal length. One is used to match the lengths of the speech signals to generate training data in the diffusion process and the other is used to generate development and inference data.
\subsection{Model Configuration}
We set the number of diffusion blocks and channels as [$N$,$C$] $\in$ [30,63],[40,128],[50,128] for small, medium, and large SEDM models (SEDM-S, SEDM-M, SEDM-L), respectively. The number of reverse blocks is equal to the number of diffusion blocks, i.e., $M=N$ .The kernel size of Bi-DilConv is 3, and the dilation is doubled at each layer within each block as [1, 2, 4, ..., $2^{n-1}$ ]. Each LSTM in CCC consists of three hidden layers and 30 features in the hidden state. Further studies on model backbones are out of scope of this paper. 

The proposed model is trained by using the Adam optimizer with a weight decay of 0.0001, a momentum of 0.9, and a batch size of 64. We train the networks for 200 epochs, where we warm-up the network in the first 20 epochs by without CCC losses. The initial learning rate is 0.03, and is multiplied by 0.1 at 120 and 160 epochs. All the experiments are run on Tesla V100 GPUs. 

Moreover, all the speech utterances are resampled to 16 kHz. They are converted to spectrogram using fast Fourier transform (FFT), with a window of 512 samples (32ms) with an overlap of 256 samples (16ms) between the neighboring windows. Since the input and the output of the proposed method and baselines are both magnitude spectrogram and the dimension of single axis is set to 257. A linear processing layer is stacked when splitting the feature map to convert the spectrogram to feature vectors of 512 dimensions.
\subsection{Competitors}
In this work, we compare the proposed method with six state-of-the-art models DCTCRN \cite{DCTCRN}, RemixIT \cite{remix}, FRCRN \cite{CCBAM}, CMGAN \cite{CMGAN}, and SCP-GAN \cite{SCP}, which reach state-of-the-art benchmarks in the DNS challenge and VCTK + DEMAND datasets. It is highlighted that we reproduce these models with the same experimental setting, e.g., training data and reverberations, as the proposed method for fair comparison.


\section{Results}
In this section, we firstly evaluate the speech enhancement performance of SEDM family and compare to state-of-the-art benchmarks on commonly used datasets, i.e., IEEE, TIMIT, VCTK, and DNS challenge. Then, we compare the proposed models to other diffusion models in the literature. Finally, we provide some visualizations and ablation study to further confirm the effectiveness of contributions.
\subsection{Evaluations on the IEEE and TIMIT Datasets}   \label{5.1}
The first experiment is conducted on IEEE and TIMIT \cite{IEEE, TM}. In the training and development stages, 600 recordings from 60 speakers and 60 recordings from 6 speakers are randomly selected in each dataset, respectively. To evaluate and compare the quality of the enhanced speech with various methods, we use the short-time objective intelligibility (STOI), perceptual evaluation of speech quality (PESQ), and frequency-weighted segmental signal-to-noise ratio (fwSNRseg) as performance measures on the IEEE and TIMIT datasets. The STOI and the PESQ are bounded in the range of [0, 1] and [-0.5, 4.5], respectively \cite{PESQ}. The fwSNRseg is estimated by computing the segmental signal-to-noise ratios (SNRs) in each spectral band and summing the weighed SNRs from all bands \cite{fw} in the range of [-10, 35] dB. 

Table 1 shows the averaged speech enhancement performance of the proposed method as compared to state-of-the-art models using the IEEE and the TIMIT datasets, with three SNR levels (-5, 0, 5 dB) and ten noise interferences in \ref{5.1}. Each result is the average of 360 noisy speech signals (120 clean speech signals $\times$3 SNR levels). From Table 1, it can be observed that in all the evaluated models, SEDM-L offers the best effectiveness.
\subsection{Evaluations on the VCTK and DNS Challenge Datasets}
We perform extensive experiments to evaluate whether SEDM family can achieve a good speech enhancement performance over the VCTK dataset. We randomly generate 11572 noisy mixtures with 10 background noises at one of 4 SNR levels (15, 10, 5, and 0 dB) in the training stage. The test set with 2 speakers, unseen during training, consists of a total of 20 different noise conditions: 5 types of noise sourced from the DEMAND dataset at one of 4 SNRs each (17.5, 12.5, 7.5, and 2.5 dB). This yields 824 test items, with approximately 20 different sentences in each condition per test speaker. To evaluate and compare the quality of the enhanced speech with various methods, we use mean opinion score (MOS) predictor of signal distortion (CSIG), MOS predictor of background intrusiveness (CBAK), MOS predictor of overall speech quality (COVL) to map the enhancement between [1, 5] \cite{PESQ}. Furthermore, similar to \cite{unet, naagn}, PESQ and segmental signal-to-noise ratio (SSNR) are used as well. Table 2 shows the averaged speech enhancement results on the VCTK dataset \cite{VB}.

\vspace{-0.5em}
\begin{table}[htbp!]
\centering
\small\addtolength{\tabcolsep}{-1pt}
\caption{Speech enhancement performance comparison on \textbf{VCTK}.}
\begin{tabular}{ccccccc}
\hline
Method   & PESQ & CSIG & CBAK & COVL & SSNR \\    
 \hline
 Unprocessed  &1.97 &3.35 &2.44 &2.63 & 1.7 \\
DCTCRN & 3.30&3.69 &3.90 & 4.53&10.1 \\
RemixIT & 3.38&3.85 &3.99 & 4.68&10.2 \\
FRCRN  &3.43 & 3.92 & 4.20&4.71 & 11.6 \\
CMGAN &3.41	& 3.94&	4.12 &4.63 & 11.1 \\
SCP-GAN & \underline{3.52} & \underline{3.97} &{\bfseries 4.25} &\underline{4.75} & 10.8 \\
\hline 
 \textit{SEDM-S} &3.46 & 3.88 & 4.07 &4.58 & 11.6\\
  \textit{SEDM-M}  &3.50 & 3.94 & 4.15 &4.71& \underline{11.7}\\
  \textit{SEDM-L} & {\bfseries3.59} &{\bfseries 4.06}&\underline{4.22}& {\bfseries 4.89} & {\bfseries 11.8}\\
 \hline 
\end{tabular}
\end{table}

From this table, we can see that the proposed method outperforms the state-of-the-art methods in terms of all performance measures. The proposed SEDM-L is 0.96 higher than SCP-GAN (11.8 vs. 10.8, SSNR).

The proposed method is further evaluated on the DNS challenge benchmark and compared with the state-of-the-art methods. The clean speech set includes over 500 hours of clips from 2150 speakers and the noise set includes over 180 hours of clips from 150 classes in the DNS challenge \cite{DNS}. In the training stage, 75$\%$ of the clean speeches are mixed with the background noise but without reverberation at a random SNR in between -5 and 20 dB as \cite{full}. In the test stage, 150 noisy clips are randomly selected from the blind test dataset without reverberations. In these experiments, the averaged STOI ($\%$), wide-band PESQ (WP), narrow-band PESQ (NP), and scale-invariant source-to-distortion ratio (SI-SDR) (dB) performances are presented in Table 3.

\vspace{-0.5em}
\begin{table}[htbp!]
\centering
\caption{Speech enhancement performance comparison on the \textbf{DNS challenge} dataset without reverberations.}
\begin{tabular}{cccccc}
\hline
Method &WP&NP&STOI ($\%$) & SI-SDR\\
 \hline
 Unprocessed  & 1.56&2.45 &91.2 & 9.0 \\
DCTCRN &2.82 &3.17 & 94.6&10.8 \\
RemixIT &{\bfseries 2.95} &\underline{3.33} & \underline{97.1}&{\bfseries 19.7} \\
FRCRN  & 2.65 & 3.23&96.1 & 11.1 \\
CMGAN & 2.54 & 3.10 &94.1 & 10.6 \\
SCP-GAN & 2.84 & 3.25 &95.2 & 10.9 \\
   \hline 
\textit{SEDM-S}  &2.88 &3.31 &96.6& 12.6\\
\textit{SEDM-M}  &2.91 & 3.35 &97.2& 13.0\\
\textit{SEDM-L}  &\underline{2.93} &{\bfseries 3.42} &{\bfseries 97.4}& \underline{13.2}\\
 \hline 
\end{tabular}
\end{table}

We observe that the proposed SEDM model shows competitive performance compared to the state-of-the-art model, Remix, on the DNS challenge. Specifically, SEDM-L outperforms the state-of-the-art models in terms of NP and STOI metrics.

\subsection{Comparison to other Diffusion Models} \label{6.3}
We further investigate the effectiveness of our diffusion model against state-of-the-art diffusion models in the literature, including denoising diffusion probabilistic model (DDPM) \cite{ddpm},  diffusion probabilistic model-based speech enhancement (DiffuSE) \cite{DM}, noise-aware speech enhancement (NASE) \cite{sedi}, score-based generative models speech enhancement (SGMSE) \cite{SGMSE}, conditional diffusion probabilistic model for speech enhancement (CDiffuSE) \cite{CDiffuse}, neural audio upsampling model (NU-Wave) \cite{NUwave}. The evaluation results on VCTK \cite{VB} + DEMAND \cite{Demand} are reported in Table 4, and the experimental setting is the same as Table 2.

\begin{table}[htbp!]
\centering
\caption{Speech enhancement performance comparison to other diffusion models on the VCTK dataset.}
\begin{tabular}{cccc}
\hline
Models & PESQ & ESTOI & SI-SDR\\
 \hline
DDPM & 2.28 &  0.64  & 8.5 \\
NU-Wave & 2.33 &  0.67  & 9.0 \\
DiffuSE & 2.41 &  0.72  & 10.9 \\
CDiffuSE & 2.58 &  0.79  & 12.4 \\
SGMSE & 2.93 &  \underline{0.87}  & 17.3 \\
NASE & \underline{3.01} &  \underline{0.87}  & \underline{17.6} \\
\textit{SEDM-L} & {\bfseries 3.59 } &{\bfseries 0.91 } &{\bfseries 19.4 }\\
\hline 
\end{tabular}
\end{table}

We observe that the proposed SEDM-L achieves the best performance, outperforming the second-best NASE \cite{sedi} by 0.58 and 0.04 over PESQ and ESTOI, respectively.
\subsection{Ablation Study}
In this experiment, we investigate the effectiveness of each contribution. A ResNet152 \cite{rc} is used when the diffusion model shows $\xmark$. Although the CCC mechanism is based on phase-aware spectrogram, we use a single diffusion model to generate speech spectra as presented in Figure 4 for the combination of (diffusion model: $\checkmark$, phase-aware: $\xmark$, and CCC: $\checkmark$).

\begin{figure}[h!]
\centering
\includegraphics[width=8cm, height=5.4cm]{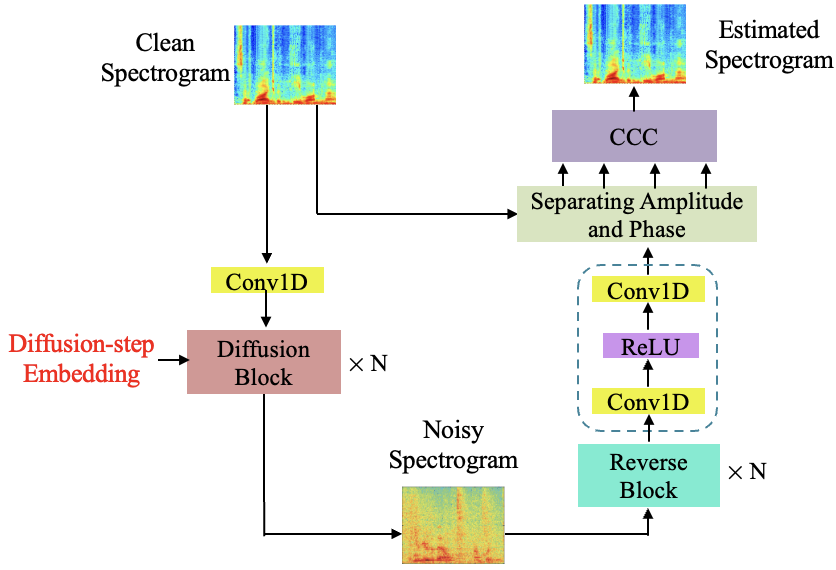}
\caption{The pipeline of (diffusion model: $\checkmark$, phase-aware: $\xmark$, and CCC: $\checkmark$). The clean speech spectra and the corresponding reconstruction are only converted into magnitude and phase components before CCC module. }\centering
\end{figure}

The models are trained and tested on the IEEE dataset as in Section \ref{5.1}. Ablation study results are showed in Table 5.

\vspace{-0.5em}
\begin{table}[htbp!]
\caption{Ablation study of the three contributions in the proposed method.}
\centering
\begin{tabular}{ccc|c}
\hline
\multicolumn{3}{c|}{Ablation Settings} & %
    \multirow{2}{*}{PESQ}   \\ 
\cline{1-3}
  Diffusion Model & Phase-Aware & CCC \\
 \hline
\xmark &\xmark &\xmark & 2.21\\
\checkmark   &\xmark &\xmark   & 2.43 \\
\xmark  &  \checkmark &\xmark  & 2.33  \\
 \hline 
\xmark  &  \checkmark &\checkmark  & 2.60\\
\checkmark   &\checkmark  &\xmark & 2.55  \\
 \hline 
\checkmark   &\checkmark   &\checkmark&{\bfseries 2.75} \\
 \hline 
\end{tabular}
\end{table}

\begin{figure*}%
\centering
\includegraphics[width=17cm, height=3.6cm]{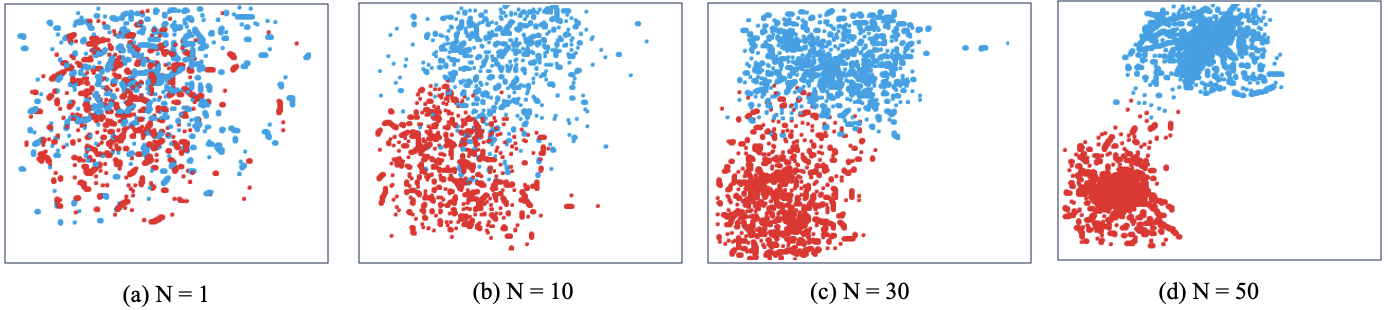}
\caption{IEEE dataset visualization with diffusion t-SNE for different numbers of embeddings N.}
\end{figure*}


Initially, we evaluate the effectiveness of diffusion model, which plays a pivotal role in learning desired features from noisy speech spectra. Diffusion model demonstrates a remarkable performance improvement from an initial PESQ of 2.21 to 2.43. This improvement can be attributed to the ability to effectively capture and model the temporal dynamics and dependencies present in speech signals. 

Moreover, speech enhancement experiences a relatively slight improvement by exploiting phase information (i.e., 2.21$\rightarrow$2.33). In the baselines, the speech signal is reconstructed by using the noisy phase and the estimated magnitude, which causes a phase loss between the clean speech signal and the corresponding reconstruction. However, the proposed phase-aware method utilizes $\theta_{A}$ and $\theta_{P}$ to estimate the phase of the target speech signal and noisy mixture, respectively, and thus improves the accuracy of estimation.

The final experiment in the ablation study involves the addition of CCC. As demonstrated in the appendix, the potential association between the magnitude and phase plays an important role in improving speech enhancement performance. With the proposed CCC mechanism, the magnitude and phase are estimated with the updated reconstruction of noisy speech features, which are better preserved in the estimated features.



\subsection{Visualization of Learned Representation}
As qualitative analysis, Figure 6 presents the t-distributed stochastic neighbour embedding (t-SNE) visualisation of the proposed model against different numbers of embeddings N using the SEDM family with C=128 and K=3 on IEEE.

Figure 6 shows the t-SNE visualisation using different perplexity settings. For small values of N, we observe that the feature embeddings are not quite separable for separation of clean speech (blue) and noise interference (red). For large perplexity values, the features representation from the SEDM-L is better separated. These t-SNE visualization results demonstrate that proposed methods are able to better learn discriminative feature representations with 50 embedding steps.
\subsection{Noise Embeddings}
In this section, we compare the proposed model trained with real-world noises to same backbones with Gaussian noise. Moreover, we evaluate the proposed model over both seen and unseen noise types in the test stage. The experimental setting is similar to Section \ref{6.3}, but we generate the test data using the remaining 5 out of 15 noise types for the unseen noise type scenario. Figure 5 presents the results.

\begin{figure}[h!]
\centering
\includegraphics[width=8.2cm, height=2.8cm]{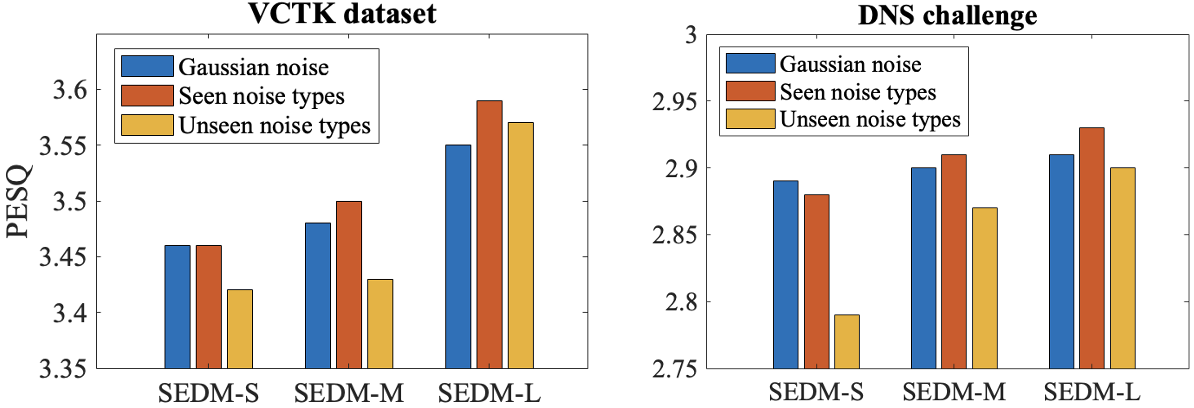}
\caption{Speech enhancement performance over VCTK dataset (left) and DNS challenge (right). The blue bars indicate models trained with Gaussian noise and evaluated with 5 of 15 noise types from DEMAND. Both red and yellow bars show models trained with real-world noise, but evaluated with seen and unseen noise types, respectively.}\centering
\end{figure}

Figure 5 shows speech enhancement performance of SEDM models against seen and unseen noise interferences in the test stage. We can observe that: (1) SEDM models trained with real-world noises suffer a performance degradation with unseen noise interferences due to noise domain mismatch. (2) SEDM-L demonstrates greater robustness compared to competitors across all noise interferences. (3) SEDM models trained with real-world noises are initially inferior to models trained with Gaussian noise in shallower networks. However, as the network depth increases, the proposed real-world noise-based models become more competitive, and in some cases, even surpass models trained with Gaussian noise.

\section{Conclusions}
In this paper, we have presented a diffusion model-based method to address the monaural speech enhancement problem. Different from the previous speech enhancement methods that ignore the intrinsic relationship between magnitude and phase information, our method estimated both the magnitude and phase information of the desired speech signal. In addition, the proposed complex-cycle-consistent mechanism provided mappings between the magnitude and phase to update the combined losses and further refined the estimation accuracy. The experimental results showed that the proposed method outperforms the state-of-the-art speech enhancement approaches over different public datasets. Our ablation experiments confirmed that real-world noise can, to a certain extent, serve as a substitute for Gaussian noise.


\section*{Acknowledgements}
This work is supported by ELSA – European Lighthouse on Secure and Safe AI funded by the European Union under grant agreement No. 101070617.

\bibliography{aaai25}

\begin{thebibliography}{38}
\providecommand{\natexlab}[1]{#1}

\bibitem[{Cao, Abdulatif, and Yang(2022)}]{CMGAN}
Cao, R.; Abdulatif, S.; and Yang, B. 2022.
\newblock {CMGAN: Conformer-based metric GAN for speech enhancement}.
\newblock \emph{IEEE International Conference on Acoustics, Speech and Signal Processing (ICASSP)}.

\bibitem[{Croitoru et~al.(2023)Croitoru, Hondru, Ionescu, and Shah}]{CVPR2}
Croitoru, F.-A.; Hondru, V.; Ionescu, R.~T.; and Shah, M. 2023.
\newblock {Diffusion models in vision: a survey}.
\newblock \emph{IEEE Transactions on Pattern Analysis and Machine Intelligence}, 45: 10850--10869.

\bibitem[{Deng et~al.(2020)Deng, Jiang, Wang, Zhang, and Li}]{naagn}
Deng, F.; Jiang, T.; Wang, X.~R.; Zhang, C.; and Li, Y. 2020.
\newblock {NAAGN: noise-aware attention-gated network for speech enhancement}.
\newblock \emph{Interspeech}.

\bibitem[{Garofolo et~al.(1993)Garofolo, Lamel, Fisher, Fiscus, Pallett, and Dahlgren}]{TM}
Garofolo, J.~S.; Lamel, L.~F.; Fisher, W.~M.; Fiscus, J.~G.; Pallett, D.~S.; and Dahlgren, N.~L. 1993.
\newblock {TIMIT acoustic phonetic continuous speech corpus CD-ROM}.
\newblock \emph{Linguistic Data Consortium}.

\bibitem[{Han and Lee(2022)}]{NUwave}
Han, S.; and Lee, J. 2022.
\newblock {NU-Wave 2: a general neural audio upsampling model for various sampling rates}.
\newblock \emph{Interspeech}.

\bibitem[{Hao et~al.(2021)Hao, Su, Horaud, and Li}]{full}
Hao, X.; Su, X.~D.; Horaud, R.; and Li, X.~F. 2021.
\newblock {FullSubNet: a full-band and sub-band fusion model for real-time single-channel speech enhancement}.
\newblock \emph{IEEE International Conference on Acoustics, Speech and Signal Processing (ICASSP)}.

\bibitem[{He et~al.(2015)He, Zhang, Ren, and Sun}]{prelu}
He, K.; Zhang, X.; Ren, S.; and Sun, J. 2015.
\newblock {Delving deep into rectifiers: surpassing human-level performance on imagenet classification}.
\newblock \emph{IEEE International Conference on Computer Vision}.

\bibitem[{He et~al.(2016)He, Zhang, Ren, and Sun}]{rc}
He, K.; Zhang, X.; Ren, S.; and Sun, J. 2016.
\newblock {Deep residual learning for image recognition}.
\newblock \emph{IEEE Conference on Computer Vision and Pattern Recognition (CVPR)}.

\bibitem[{Hu et~al.(2020)Hu, Chen, Li, Zhu, and Chng}]{ddpm}
Hu, Y.; Chen, C.; Li, R.; Zhu, Q.; and Chng, E.~S. 2020.
\newblock {Denoising diffusion probabilistic models}.
\newblock \emph{Processdings of Neural Information Processing Systems (NeurIPS)}.

\bibitem[{Hu et~al.(2023)Hu, Chen, Li, Zhu, and Chng}]{sedi}
Hu, Y.; Chen, C.; Li, R.; Zhu, Q.; and Chng, E.~S. 2023.
\newblock {Noise-aware speech enhancement using diffusion probabilistic model}.
\newblock \emph{arXiv preprint arXiv:2307.08029}.

\bibitem[{Hu and Loizou(2008)}]{PESQ}
Hu, Y.; and Loizou, P.~C. 2008.
\newblock {Evaluation of objective quality measures for speech enhancement}.
\newblock \emph{IEEE Transactions on Audio, Speech and Language Processing}, 16(1): 229--238.

\bibitem[{{IEEE Audio and Electroacoustics Group}(1969)}]{IEEE}
{IEEE Audio and Electroacoustics Group}. 1969.
\newblock {IEEE recommended practice for speech quality measurements}.
\newblock \emph{IEEE Transactions on Audio, Speech and Language Processing}, AE-17(3): 225--246.

\bibitem[{Li et~al.(2021{\natexlab{a}})Li, Gao, Guan, and Ma}]{DCTCRN}
Li, Q.; Gao, F.; Guan, H.; and Ma, K. 2021{\natexlab{a}}.
\newblock {Real-time monaural speech enhancement with short-time discrete cosine transform}.
\newblock \emph{arXiv preprint arXiv: 2102.04629}.

\bibitem[{Li et~al.(2021{\natexlab{b}})Li, Sun, Horoshenkov, and Naqvi}]{TAI}
Li, Y.; Sun, Y.; Horoshenkov, K.; and Naqvi, S.~M. 2021{\natexlab{b}}.
\newblock {Domain adaptation and autoencoder based unsupervised speech enhancement}.
\newblock \emph{IEEE Transactions on Artificial Intelligence}, 3(1): 43 -- 52.

\bibitem[{Li et~al.(2023)Li, Sun, Wang, and Naqvi}]{ushaped}
Li, Y.; Sun, Y.; Wang, W.; and Naqvi, S.~M. 2023.
\newblock {U-shaped Transformer with frequency-band aware attention for speech enhancement}.
\newblock \emph{IEEE/ACM Transactions on Audio, Speech and Language Processing}, 31: 1511–1521.

\bibitem[{Liu, Ma, and Chen(2017)}]{fw}
Liu, Z.~X.; Ma, H.~T.; and Chen, F. 2017.
\newblock {A new data-driven band-weighting function for predicting the intelligibility of noise-suppressed speech}.
\newblock \emph{Asia-Pacific Signal and Information Processing Association Annual Summit and Conference (APSIPA ASC)}.

\bibitem[{Lu, Tsao, and Watanabe(2021)}]{DM}
Lu, Y.-J.; Tsao, Y.; and Watanabe, S. 2021.
\newblock {A study on speech enhancement based on diffusion probabilistic model}.
\newblock \emph{Asia-Pacific Signal and Information Processing Association Annual Summit and Conference (APSIPA ASC)}.

\bibitem[{Lu et~al.(2022)Lu, Wang, Watanabe, Richard, Yu, and Tsao}]{CDiffuse}
Lu, Y.-J.; Wang, Z.-Q.; Watanabe, S.; Richard, A.; Yu, C.; and Tsao, Y. 2022.
\newblock {Conditional diffusion probabilistic model for speech enhancement}.
\newblock \emph{IEEE International Conference on Acoustics, Speech and Signal Processing (ICASSP)}.

\bibitem[{Macartney and Weyde(2018)}]{unet}
Macartney, C.; and Weyde, T. 2018.
\newblock {Improved speech enhancement with the wave-u-net}.
\newblock \emph{arXiv preprint arXiv:1811.11307}.

\bibitem[{Rahman, J.~M. J.~Valanarasu, and Patel(2023)}]{CVPR1}
Rahman, A.; J.~M. J.~Valanarasu, I.~H.; and Patel, V.~M. 2023.
\newblock {Ambiguous medical image segmentation using diffusion models}.
\newblock \emph{IEEE Conference on Computer Vision and Pattern Recognition (CVPR)}.

\bibitem[{Reddy et~al.(2021)Reddy, Dubey, Koishida, Nair, Gopal, Cutler, Braun, Gamper, Aichner, and Srinivasan}]{DNS}
Reddy, C.; Dubey, H.; Koishida, K.; Nair, A.; Gopal, V.; Cutler, R.; Braun, S.; Gamper, H.; Aichner, R.; and Srinivasan, S. 2021.
\newblock {Interspeech 2021 deep noise suppression challenge}.
\newblock \emph{Interspeech}.

\bibitem[{Rogers and Williams(2000)}]{sde}
Rogers, L. C.~G.; and Williams, D. 2000.
\newblock {Diffusions, Markov processes and martingales,}.
\newblock \emph{Cambridge University Press}.

\bibitem[{Shimauchlt et~al.(2017)Shimauchlt, Kudo, Koizumli, and Furuva}]{phase}
Shimauchlt, S.; Kudo, S.; Koizumli, Y.; and Furuva, K. 2017.
\newblock {On relationships between amplitude and phase of short-time fourier transform}.
\newblock \emph{IEEE International Conference on Acoustics, Speech and Signal Processing (ICASSP)}.

\bibitem[{Sohl-Dickstein et~al.(2015)Sohl-Dickstein, Weiss, Maheswaranathan, and Ganguli}]{DM0}
Sohl-Dickstein, J.; Weiss, E.; Maheswaranathan, N.; and Ganguli, S. 2015.
\newblock Deep unsupervised learning using nonequilibrium thermodynamics.
\newblock \emph{International Conference on Machine Learning (ICML)}.

\bibitem[{Thiemann, Ito, and Vincent(2013)}]{Demand}
Thiemann, J.; Ito, N.; and Vincent, E. 2013.
\newblock {The diverse environments multi-channel acoustic noise database: a database of multichannel environmental noise recordings}.
\newblock \emph{The Journal of the Acoustical Society of America}, 133(5): 3591 -- 3591.

\bibitem[{Tzinis et~al.(2022)Tzinis, Adi, Ithapu, Xu, Smaragdis, and Kumar}]{remix}
Tzinis, E.; Adi, Y.; Ithapu, V.~K.; Xu, B.; Smaragdis, P.; and Kumar, A. 2022.
\newblock {RemixIT: continual self-training of speech enhancement models via bootstrapped remixing}.
\newblock \emph{IEEE Journal of Selected Topics in Signal Processing}, 16(6): 1329--1341.

\bibitem[{Veaux, Yamagishi, and King(2013)}]{VB}
Veaux, C.; Yamagishi, J.; and King, S. 2013.
\newblock {The voice bank corpus: design, collection and data analysis of a large regional accent speech database}.
\newblock \emph{IEEE Conference on Asian Spoken Language Research and Evaluation (O-COCOSDA/CASLRE)}.

\bibitem[{Wang, Narayanan, and Wang(2014)}]{IRM1}
Wang, Y.; Narayanan, A.; and Wang, D. 2014.
\newblock {On training targets for supervised speech separation}.
\newblock \emph{IEEE/ACM Transactions on Audio, Speech and Language Processing}, 22(12): 1849--1858.

\bibitem[{Welker, Richter, and Gerkmann(2022{\natexlab{a}})}]{GN}
Welker, S.; Richter, J.; and Gerkmann, T. 2022{\natexlab{a}}.
\newblock {Speech enhancement with score-based generative models in the complex STFT domain}.
\newblock \emph{Interspeech}.

\bibitem[{Welker, Richter, and Gerkmann(2022{\natexlab{b}})}]{SGMSE}
Welker, S.; Richter, J.; and Gerkmann, T. 2022{\natexlab{b}}.
\newblock {Speech enhancement with score-based generative models in the complex STFT domain}.
\newblock \emph{Interspeech}.

\bibitem[{Williamson, Wang, and Wang(2016)}]{comp}
Williamson, D.~S.; Wang, Y.; and Wang, D. 2016.
\newblock Complex ratio masking for monaural speech separation.
\newblock \emph{IEEE/ACM Transactions on Audio, Speech and Language Processing}, 24(3): 483 -- 492.

\bibitem[{Wu et~al.(2023)Wu, Han, Jiang, Fan, Zeng, and Liu}]{realnoise}
Wu, Q.; Han, M.; Jiang, T.; Fan, H.; Zeng, B.; and Liu, S. 2023.
\newblock {Realistic noise synthesis with diffusion models}.
\newblock \emph{arXiv preprint arXiv: 2305.14022}.

\bibitem[{Yang, Pandey, and Wang(2023)}]{seasr1}
Yang, Y.; Pandey, A.; and Wang, D. 2023.
\newblock {Time-domain speech enhancement for robust automatic speech recognition}.
\newblock \emph{Interspeech}.

\bibitem[{Yu et~al.(2021)Yu, Zhou, Wang, and Tao}]{setf}
Yu, W.~W.; Zhou, J.; Wang, H.~B.; and Tao, L. 2021.
\newblock {SETransformer: speech enhancement transformer}.
\newblock \emph{Cognitive Computation}.

\bibitem[{Zadorozhnyy and Q.~Ye(2022)}]{SCP}
Zadorozhnyy, V.; and Q.~Ye, K.~K. 2022.
\newblock {SCP-GAN: self-correcting discriminator optimization for training consistency preserving metric GAN on speech enhancement tasks}.
\newblock \emph{arXiv preprint arXiv: 2210.14474}.

\bibitem[{Zhao, Nguyen, and Ma(2021)}]{CCBAM}
Zhao, S.; Nguyen, T.~H.; and Ma, B. 2021.
\newblock {Monaural speech enhancement with complex convolutional block attention module and joint time frequency losses}.
\newblock \emph{IEEE International Conference on Acoustics, Speech and Signal Processing (ICASSP)}.

\bibitem[{Zhao and Wang(2020)}]{tf3}
Zhao, Y.; and Wang, D.~L. 2020.
\newblock {Noisy-reverberant Speech Enhancement Using DenseUNet with Time-frequency Attention}.
\newblock \emph{Interspeech}.

\bibitem[{Zhu et~al.(2023)Zhu, Zhang, Zhang, and Dai}]{seasr}
Zhu, Q.-S.; Zhang, J.; Zhang, Z.-Q.; and Dai, L.-R. 2023.
\newblock {A joint speech enhancement and self-supervised representation learning framework for noise-robust speech recognition}.
\newblock \emph{IEEE/ACM Transactions on Audio, Speech and Language Processing}, 31: 1927--1939.

\end{thebibliography}
\clearpage
\appendix
\section{Quantitative Results}
\subsection{Intrinsic Relationship between Magnitude and Phase} \label{ap}
In the proposed method, the CCC mechanism is used to minimize the differences between the reconstructed magnitude/phase and the corresponding magnitude/phase from clean speech signals. We also conduct experiments to evaluate the CCC mechanism. The averaged experimental results on IEEE and TIMIT as in Section 6.1 are shown in Table 1.

\begin{table}[htbp!]
\centering
\small\addtolength{\tabcolsep}{-1pt}
\setlength{\tabcolsep}{3.2pt}
\caption{Speech enhancement performance by estimating the magnitude and phase by using their losses. The magnitude and phase are represented as A and P, respectively.}
\centering
\begin{tabular}{c|ccc}
\hline
& STOI ($\%$)  & PESQ & fwSNRseg (dB) \\
\hline
Noisy A + Noisy P   & 42.3 & 1.52 & 3.11 \\

Estimated A with P loss  & 49.4 & 1.68 & 5.62  \\

Estimated P with A loss  & 44.6 & 1.59 & 3.75 \\  
\hline
\begin{tabular}[c]{@{}l@{}}Estimated A with P loss +\\ Estimated P with A loss\end{tabular} & 50.7 & 1.71 & 5.91   \\ 
\hline
\end{tabular}
\end{table}
Table 1 shows that relationship between magnitude and phase, as shown in \cite{phase}, is beneficial for further enhancing speech enhancement performance. This seems to be aligned with the findings in \cite{phase} which shows that the magnitude and phase can be linked through the group delay or instantaneous frequency without making any assumptions on the phase property of the target signals, e.g., minimum, maximum, or linear phase.
\subsection{Benckmarks on TIMIT.}
Furthermore, we compare the proposed models to the benchmarks of all competitor models on TIMIT \cite{TM}. The results are shown in Table 2.

\begin{table}[ht]
\centering
\caption{Comparison on TIMIT.}
\begin{tabular}{lccc}
\toprule
Model     & STOI & PESQ & fwSNRseg (dB) \\
\midrule
DCTCRN    & 78.5 & 2.45 & 13.2     \\
RemixIT   & 79.3 & 2.56 & 14.0     \\
FRCRN     & 80.2 & 2.59 & 14.4     \\
CMGAN     & 79.6 & 2.55 & 13.8     \\
SCP-GAN   & 81.5 & 2.77 & 15.0     \\
DDPM      & 73.6 & 2.20 & 11.1     \\
NU-Wave   & 74.8 & 2.35 & 11.8     \\
DiffuSE   & 74.7 & 2.32 & 12.2     \\
CDiffuSE  & 78.6 & 2.53 & 13.5     \\
SGMSE     & 77.4 & 2.49 & 12.9     \\
NASE      & 81.0 & 2.69 & 14.6     \\
SEDM-S    & 81.2 & 2.71 & 14.9     \\
SEDM-M    & \underline{81.7} & \underline{2.81} & \underline{15.2}     \\
SEDM-L    & \textbf{81.9} & \textbf{2.83} & \textbf{15.3}     \\
\bottomrule
\end{tabular}
\end{table}

From Table 2, SEDM-L demonstrates the best overall performance across the metrics, with the highest STOI of 81.9, PESQ of 2.83, and fwSNRseg of 15.3. This indicates that SEDM-L excels in improving speech enhancement performance compared to other models.
\section{Qualitative Results}
In this section, we present qualitative results demonstrating the enhancement performance of noisy speech samples. Particularly, we show the qualitative improvements of estimated signal with our SEDM-L over noisy signal and SCP-GAN \cite{setf} on TIMIT in Figure 1.

\begin{figure}[h!]
\centering
\includegraphics[width=8.2cm, height=5.6cm]{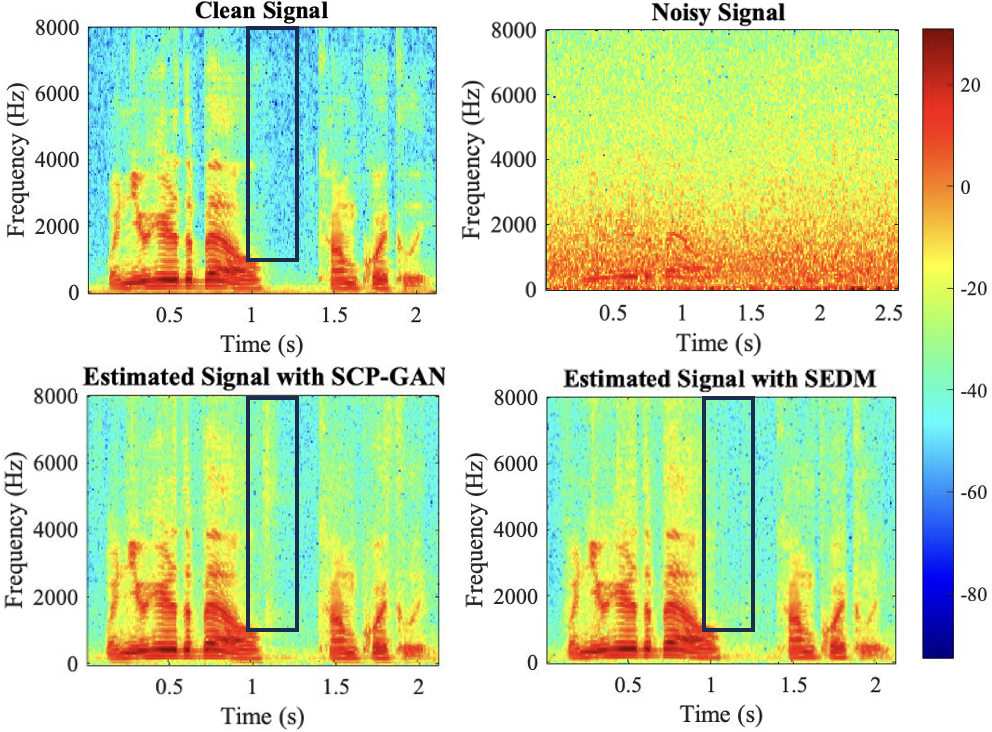}
\caption{The spectra of different signals. The experiment is implemented with $driver$ and -5 dB SNR level. The proposed SEDM model offers 0.11 and 0.20 improvements over the best-performing baseline, i.e. SCP-GAN \cite{setf}, in terms of PESQ and STOI, respectively.}\centering
\end{figure}
After comparing the estimated spectra with the spectrogram of target speech signal, it can be observed that the spectrogram obtained via the proposed SEDM-L is closer to the clean speech signal, which again confirms that the proposed method outperforms the competitors.
\end{document}